\documentclass[prc,aps,showpacs,twocolumn,floatfix]{revtex4}
\usepackage{graphicx}

\begin{document}

\title{Counter Terms for Low Momentum Nucleon-Nucleon Interactions } 
\author{ Jason D. Holt $^{1}$, T.T.S. Kuo$^{1}$, G.E. Brown $^{1}$
and Scott K. Bogner $^{2}$ }
\affiliation{ $^1$Department of Physics, State University of New York at Stony
Brook Stony Brook, New York 11794\\
$^2$ Institute for Nuclear Theory, Univ. of Washington, Seattle, WA 96195 
}
\date{\today}

\begin{abstract}
There is much current interest in treating low energy nuclear physics 
using the renormalization group (RG) and effective field theory (EFT). 
Inspired by 
this RG-EFT approach, we study  a low-momentum  nucleon-nucleon (NN) 
interaction, $V_{low-k}$,  obtained by integrating out the fast modes down 
to the scale $\Lambda \sim 2 \rm {fm} ^ {-1}$. Since NN experiments can only 
determine the effective interaction in this low momentum region, our chief 
purpose is to find such an interaction for complex nuclei whose typical 
momenta lie below this scale. In this paper we find that $V_{low-k}$ can be 
highly satisfactorily accounted for by the counter terms 
corresponding to a short range effective interaction. 
The coefficients $C_n$ of the power series expansion $\sum C_n q^n$
for the counter terms  have been accurately determined, and results 
derived from several meson-exchange NN interaction models are compared. 
The counter terms are  found to be important only for the S, P and D
partial waves. Scaling behavior of the counter terms is studied.
Finally we discuss the use of these methods for computing shell 
model matrix elements.
\end{abstract}

\maketitle

\section{Introduction}

Since the pioneering work of Weinberg \cite{weinberg}, there has been much 
progress and interest in treating low-energy nuclear physics using the 
renormalization group (RG) and effective field theory (EFT) approach 
\cite{meissner,lepage,kaplan98,epel99,EFT,EFT02,EFT03,kolck,haxton,beane}. 
A central idea here is that 
physics in the infra red region must be insensitive to the details of the 
short range (high momentum) dynamics. In low-energy nuclear physics, we 
are 
probing nuclear systems with low-energy probes of wave length $\lambda$; 
such 
probes cannot reveal the short range details at distances much 
smaller than $\lambda$. Furthermore, our understanding about the short 
range 
dynamics is still preliminary, and model dependent. Because of these 
considerations, a central step in the RG-EFT approach is to divide the 
fields 
into two categories: slow fields and fast fields, separated by a chiral 
symmetry breaking scale $\Lambda_{\chi} \sim$ 1 GeV.  Then by integrating 
out 
the fast fields, one obtains an effective field theory for the slow fields 
only. This approach has been very successful in treating low-energy 
nuclear 
systems, as discussed in the references cited above. 

In order to have an 
effective interaction appropriate for complex nuclei in which typical 
nucleon momenta are $< k_F$, we separate fast and
slow modes by a scale much smaller than $\Lambda_{\chi}$,
namely $\Lambda \sim 2~ \rm{fm^{-1}}$. This corresponds to the pion production 
threshold, a scale
often not considered in traditional EFT methods, even though it represents the
limit of available 2-nucleon data; experiments give
a unique effective interaction only up to this scale. This cutoff has been 
employed recently by several authors \cite{bogner01,bogner02,bogner03,bogner04,bogner05,kuorg02,schwenk02,coraggio02,sncoraggio}
in a new development for the nucleon-nucleon (NN) 
interaction: the derivation of a low momentum NN potential, $V_{low-k}$. 
Despite some general similarities, we would like to stress that $V_{low-k}$ 
is not a traditional EFT construct, and would like to now attempt to clarify
this issue.

One drawback to the commonly used RG-EFT, which adheres to order-by-order 
consistency in the power counting, is it's lack of predictability.  The number 
of unknown parameters increases rapidly as the number of nucleons involved
in the process increases. Even if such a method were available for complex 
nuclei, the parameters would be far from first principle QCD. 
% Our approach, developed in \cite{bogner01,bogner02,bogner03,bogner04,bogner05,kuorg02,schwenk02,coraggio02,sncoraggio}, is admittedly indirect and drastically simplified. 
The key ingredient in our approach is the
highly refined standard nuclear physics approach (SNPA) and the objective 
is to marry the SNPA to an EFT \cite{brownrho02}.  

As an illustration of our strategy, we briefly discuss how this marriage can 
be effectuated. As summarized recently \cite{park,brownrho}, a thesis now 
develped for some
time posits that by combining the SNPA, based on potentials fit to experiments,
with modern effective field theory, one can achieve more predictive power than
the purist's EFT alone.
This combination of SNPA and effective field theory, called EFT* by Kubodera 
\cite{kubodera}, but more properly called {\it more effective} effective field
theory (MEEFT) as noted by Kubodera, was rediscovered in 
\cite{bogner01,bogner02,bogner03,bogner04,bogner05,kuorg02,schwenk02,coraggio02,sncoraggio},
and consists of
an RG-type approach which limits $\Lambda$ to 2.1$\rm fm^{-1}$, because this 
is equal to the center of mass momentum up to which experiments measuring
the nucleon-nucleon scattering phase shifts have been carried out.

Thus, the ``more effective'' in MEEFT arises from the guarantee that all 
experimental data will be reproduced by $V_{low-k}$, which will contain no 
Fourier components higher than those at which experiments have been carried 
out and analyzed.  Therefore, $V_{low-k}$ is reliable up to $k=\Lambda$.  
This seems to be adequate for a description of shell model properties of 
complex nuclei.

In practice, the construction of $V_{low-k}$ begins with one of a number 
of available realistic models 
\cite{cdbonn,nijmegen,argonne,chiralvnn,paris} for the 
nucleon-nucleon (NN) potential $V_{NN}$. While these models agree well in 
the low momentum (long range) region, where they are just given by the one 
pion 
exchange interaction, in the high momentum (short range) region they are 
rather uncertain and, in fact, differ significantly from one another. 
Naturally, it would be desirable to remove these uncertainties and model 
dependence
from the high momentum components of the various modern NN potentials.
Following our RG-type approach to achieve this end, it would seem to be 
appropriate
to integrate out the high momentum modes of the various $V_{NN}$ models.
This is how the $V_{low-k}$ was derived, and in section II, we provide a 
brief
outline of the derivation based on T-matrix equivalence.

The low-momentum NN potential, $V_{low-k}$, reproduces the deuteron 
binding 
energy, low-energy NN phase shifts and the low-momentum half-on-shell 
T-matrix. Furthermore, it is a smooth potential and can be used directly 
in 
nuclear many body calculations, avoiding the calculation of the Brueckner 
G-matrix \cite{bogner01,bogner02,kuorg02,schwenk02,coraggio02,sncoraggio}. 
Shell model nuclear structure calculations using $V_{low-k}$
have indeed yielded very encouraging results 
\cite{bogner02,coraggio02,sncoraggio}. These calculations apply the same
$V_{low-k}$ to a wide range of nuclei, including those in the sd-shell,
tin region, and heavy nuclei in the lead region.
At the present, no algebraic form exists for $V_{low-k}$, and 
it should be useful to have one. In the present work we 
study the feasibility of finding such an expression.

A central result of modern renormalization theory is that a general
RG decimation generates an infinite series of counter terms 
\cite{lepage,kaplan98,epel99,EFT,kolck} 
consistent with the input interaction. When we derive our
low momentum interaction, the high momentum modes of the input interaction
are integrated out. Does this decimation also generate a series of counter 
terms?  If so, then what are the properties of the counter 
terms so generated? We study these 
questions in section III where we carry out an accurate determination of 
the 
counter terms and show that this approach reproduces not only $V_{low-k}$, 
but
the deuteron binding energy and low-energy phase shifts as well. 
We shall discuss that the counter terms represent generally a short
range effective interaction and are important only for partial waves with
angular momentum $l\leq 2$. The scaling behavior of the counter terms
with respect to the decimation momentum will be studied in section IV.
Finally in section V, we examine the prospect for using $V_{low-k}$ and the 
counter term method in shell model calculations.

%We shall also
% compare the Sussex shell-model matrix elements of  
%Elliott et al. \cite{elliott} with those given by our $V_{low-k}$;
%the Sussex shell-model
%matrix elements were based on low-energy NN phase shifts.

%\section{$V_{low-k}$, EFT, and MEEFT}

\section{T-matrix equivalence}

Since our method for deriving the low-momentum interaction $V_{low-k}$
has been described elsewhere
\cite{bogner01,bogner02,kuorg02}, in the following we only outline 
the derivation.  We obtain  $V_{low-k}$ from a realistic $V_{NN}$ model, 
such as the CD-Bonn model, by integrating out the  high-momentum 
components.
This integration is carried out with the requirement that the deuteron 
binding
energy and low-energy phase shifts of $V_{NN}$ are preserved by 
$V_{low-k}$. 
This preservation may be satisfied by the following T-matrix equivalence 
approach \cite{bogner01,bogner02,kuorg02}.
We start from the half-on-shell T-matrix
\begin{eqnarray}
  T(k',k,k^2)  = V_{NN}(k',k)   + \int _0 ^{\infty} q^2 dq  V_{NN}(k',q) 
\nonumber \\
  \times \frac{1}{k^2-q^2 +i0^+ } T(q,k,k^2 ) ,
\end{eqnarray}
noting that the intermediate state momentum $q$ is integrated from 0 to 
$\infty$.
We then define an effective low-momentum T-matrix by 
\begin{eqnarray}
  T_{low-k }(p',p,p^2) & = & V_{low-k }(p',p)
  + \int _0 ^{\Lambda} q^2 dq  V_{low-k }(p',q) \nonumber \\
& & \times \frac{1}{p^2-q^2 +i0^+ } T_{low-k} (q,p,p^2)
\end{eqnarray}
where $\Lambda$ denotes a momentum space cut-off and $(p',p)\leq \Lambda$. 
We choose $\Lambda\sim \rm{2fm^{-1}}$, essentially the momentum up to which the
experiments give us information in the phase shift analysis. Note that
in Eq.(2) the intermediate 
state momentum is integrated from 0 to $\Lambda$. We require the above 
T-matrices satisfying the condition
\begin{equation}
 T(p',p,p^2 ) = T_{low-k }(p',p, p^2 ) ;~( p',p) \leq \Lambda.
\end{equation}
The above equations define  the effective low momentum interaction
 $V_{low-k}$, and are satisfied by the solution 
\cite{bogner01,bogner02,kuorg02}   
\begin{eqnarray}
  V_{low-k} &= & \hat{Q} - \hat{Q'} \int \hat{Q} + \hat{Q'} \int \hat{Q} \int 
  \hat{Q}  \nonumber \\
   && {} - \hat{Q'} \int \hat{Q} \int \hat{Q} \int \hat{Q} + \cdots 
\end{eqnarray}
which is  just the Kuo-Lee-Ratcliff 
folded-diagram effective interaction \cite{klr71,ko90}.
Here $\hat Q$ represents the irreducible vertex function whose 
intermediate
states are all beyond $\Lambda$; $\hat {Q'}$ is the same as $\hat Q$
except with its terms first order in the interaction removed. 

For any decimation momentum $\Lambda$, the above $V_{low-k}$ can be 
calculated highly accurately (essentially exactly) using either the 
Andreozzi-Lee-Suzuki 
\cite{suzuki80,andre96} (ALS) or the Krenciglowa-Kuo \cite{krmku74} 
iteration methods. These procedures  preserve the deuteron binding energy 
in 
addition to the half-on-shell T matrix, which 
implies preservation of the phase shifts. After obtaining the above
$V_{low-k}$, which is not Hermitian, we further perform an Okubo 
transformation
\cite{okubo} to make it Hermitian. This is done using the method given
in Refs. \cite{suzuki83,kuo93}. Here we first calculate the eigenvalues
and eigenvectors of the operator $\omega ^{\dagger} \omega$ where $\omega$
is the wave operator obtained with the ALS method. Then the Hermitian
$V_{low-k}$ is calculated in terms of these quantities in a convenient
way (see Eq.(23) of Ref.\cite{kuo93}). 

\section{Counter terms}

\begin{figure}
\rotatebox{270}{\scalebox{0.4}{\includegraphics{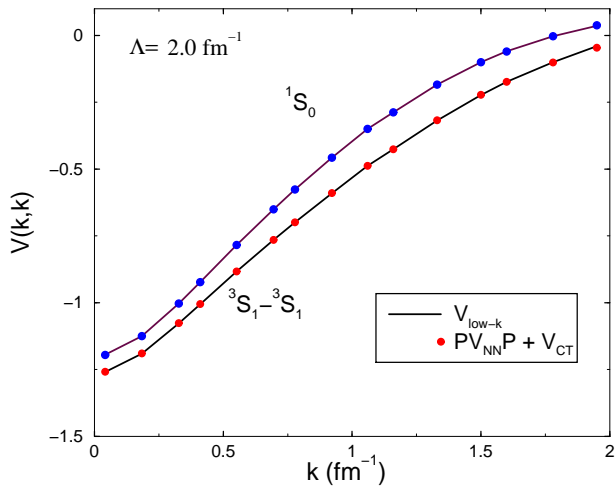}}}
\caption{Comparison of $V_{low-k}$ with $PV_{NN}P$ plus counter terms,
for $^1S_0$ and $^3S_1$ channels.}
\label{fig.1}
\end{figure}

\begin{figure}
\rotatebox{270}{\scalebox{0.4}{\includegraphics{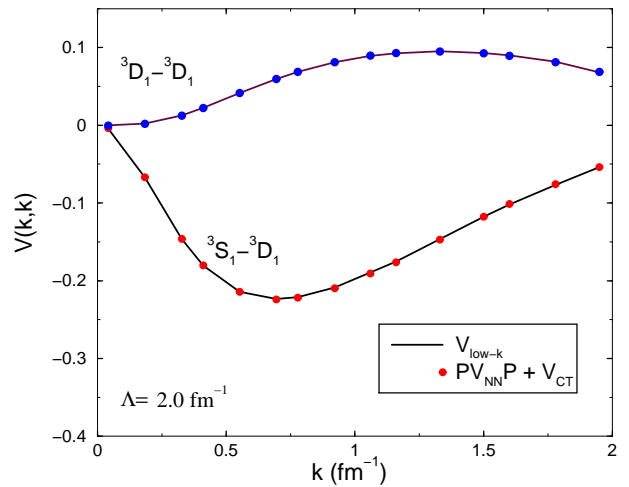}}}
%{\epsfig{file=ctholtfig2.eps,width=14cm,angle=270}}
\caption{Comparison of $V_{low-k}$ with $PV_{NN}P$ plus counter terms,
for $^3S_1-$$^3D_1$ and $^3D_1$ channels.}
\label{fig.2}
\end{figure}

Here we study whether the low-momentum interaction $V_{low-k}$ can be 
well represented by the low-momentum part of the original
NN interaction supplemented by certain simple counter terms. Specifically,
we consider
\begin{equation}
V_{low-k}(q,q')\simeq V_{bare}(q,q')+V_{counter}(q,q'); 
~(q,q')\leq \Lambda
\end{equation}
where $V_{bare}$ is some bare NN potential $V_{NN}$. 
Since $V_{low-k}$ and $V_{bare}$ are both known, 
there is no question about the existence  of  $V_{counter}$ in general.
 Our aim is, however, to investigate
if $V_{counter}$ can be well represented by  a short ranged effective 
interaction, 
such as a smeared out delta function. We shall check this conjecture
by assuming a suitable momentum expansion form for $V_{counter}$ and
investigating how well it can satisfy the above equality.
Since $V_{bare}$ is generally given according to partial waves as is
$V_{low-k}$, we shall proceed to determine $V_{counter}$ separately
for each partial wave. At very small radial distance, the Bessel function
$j_l(qr)$ behaves like $(qr)^l$. We assume that $V_{counter}$ is a very short
ranged interaction. Hence the leading term in a momentum expansion
of the partial wave matrix element $\langle ql | V_{counter} | q'l' \rangle$
is proportional to $q^l q'^{l'}$. Thus we consider the following expansion
for the partial wave counter term potential, namely
\vskip .1cm
\begin{eqnarray}
\langle ql |V_{counter}|q' l'\rangle &=&
  q^l q'^{l'}[ C_0+C_2(q^2+q'^2) 
     \nonumber \\
&&   +C_4(q^2+q'^2)^2+ C_6(q^6+q'^6) \nonumber \\
&&   +C'_4 q^2q'^2+C_6'q^4q'^2+C_6''q^2q'^4 +\cdots] \nonumber \\ 
\end{eqnarray}
The counter term coefficients will be determined by standard $\chi $-square
fitting procedure so that the difference between $V_{low-k}$
and ($V_{bare}$+$V_{counter}$) is minimized. 

\begin{figure}
\rotatebox{270}{\scalebox{0.4}{\includegraphics{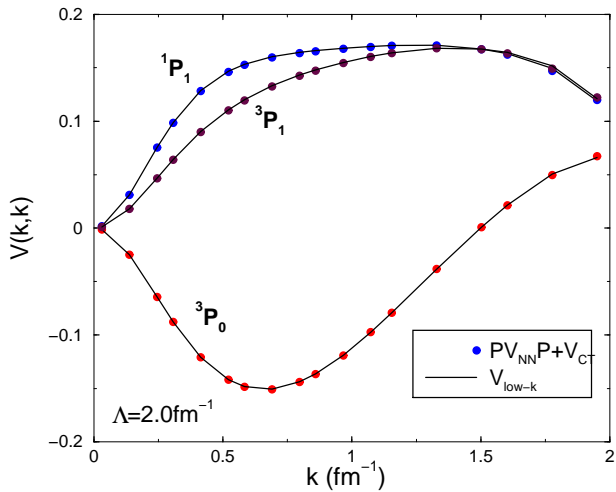}}}
%{\epsfig{file=ctholtfig2.eps,width=14cm,angle=270}}
\caption{Comparison of $V_{low-k}$ with $PV_{NN}P$ plus counter terms,
for P-wave channels.}
\label{fig.3}
\end{figure}

We note that for $l$ and $l'$ being both S-wave, the 
leading term for $V_{counter}$ is $C_0$ which is momentum independent
and corresponds to a delta interaction. For other channels, however,
the leading term for the counter potential is momentum dependent. For example,
for $l$ and $l'$ being both P-wave, the leading term  is $qq'C_0$.

If we take the projection
operator $P$ to project onto states with momentum less than $\Lambda$,
namely $P=\sum _{k<\Lambda}\mid k \rangle \langle k \mid$, we can rewrite
$V_{bare}$ of Eq.(5) as $PV_{bare}P$. 
From now on we shall also abbreviate $V_{counter}$ as $V_{CT}$.
The above counter term approach for $V_{low-k}$ is similar to that used 
in RG-EFT \cite{lepage,epel99}, as both methods impose some cutoff,
integrate out the modes above that cutoff, and then attempt to recover the 
information contained in those states in a counter term series.
Ultimately, the main difference is the the cutoff scale used and the resulting
potential to which $V_{CT}$
is added. As already noted, in the representation of 
$V_{low-k}$ given above, we use a momentum cutoff above which no constraining 
data exists, and thus our ``slow'' potential is a bare NN potential projected
onto the low momentum states. In the EFT scheme, the ``slow'' potential is 
obtained by integrating out heavy dynamical degrees of freedom above the 
chiral symmetry breaking
scale. Hence, counter terms are added
to $V_{light}$, which is given by light mesons below the cutoff scale, and 
usually taken to be $V_{one-pion}$ with unrestricted pion momentum.

We have also tried to fit $V_{low-k}$ solely to a low order momentum expansion,
and found the results to be less than satisfactory; the number of terms needed 
to produce an accurate representation is much higher than the order used here. 
%This is to be expected, however, as 
%a low order expansion corresponds to a short-range interaction, 
%and $V_{low-k}$ is a long-range interaction.  ???Future work???
Attempts at fitting ($V_{one-pion}$+ $V_{CT}$) also generated results 
far less satisfactory than fitting it by ($V_{bare}$+$V_{CT}$).  
We shall discuss these points later.

\begin{figure}
\rotatebox{270}{\scalebox{0.4}{\includegraphics{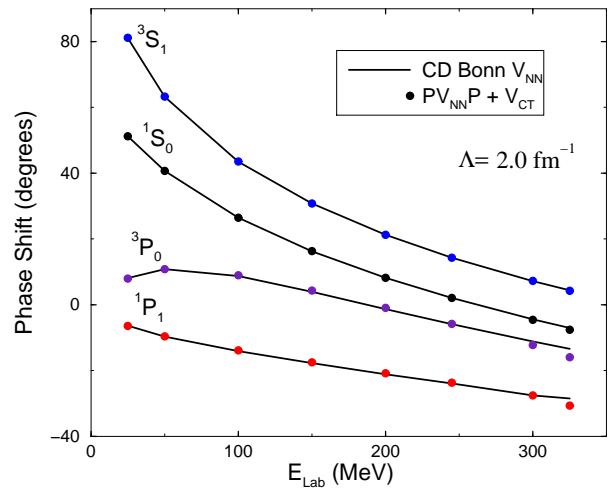}}}
%\centerline {\epsfig{file=ctholtfig3.eps,width=14cm,angle=270}}
\caption{Comparison of phase shifts given by $V_{NN}$ 
and  $PV_{NN}P $ plus counter terms.}
\label{fig.4}
\end{figure}

\begin{table*}
\caption{Listing of counter terms for all partial waves obtained 
from the CD-Bonn potential using $\Lambda = \rm{2fm^{-1}}$. The units for the 
combined quantity $q^lq'^{l'}C_nq^n$ of Eq.(6) is fm, with momentum $q$ 
in $\rm{fm}^{-1}$. All counter terms for higher order partial 
waves are zero up to 
our numerical accuracy.}
\vskip 0.5cm
\begin{tabular}{ccccccccc}

Wave       &  $C_0$  &   $C_2 $  &   $C_4$   & $C_4' $  &   $C_6$  
	                        &   $C_6'$  &   $C_6''$ & $\Delta_{rms}$\\    
\hline 
$^{1}S_{0}$           & -0.1580   & -1.309E-2 &  3.561E-4 & -8.469E-4 
                      &  0        & -1.191E-4 & -1.191E-4 &  0.0002    \\
$^{3}S_{1}$           & -0.4651   &  5.884E-2 & -1.824E-3 & -1.163E-2 
                      & -3.243E-4 &  5.181E-4 &  5.181E-4 &  0.0003    \\
$^{3}S_{1}-^{3}D_{1}$ &  2.879E-2 & -3.581E-3 &  1.969E-3 & -3.676E-3 
                      & -1.376E-4 &  6.300E-4 & -2.449E-4 &  0.0025    \\
$^{3}D_{1}$           & -1.943E-3 & -1.918E-4 &  1.709E-4 &  0     
                      &  0        &  0        &  0        & $<$0.0001  \\
$^{1}P_{1}$           & -4.311E-2 & -7.305E-4 &  7.999E-4 & -2.049E-4 
                      & -2.594E-4 &  0        &  0        &  0.0002    \\
$^{3}P_{0}$           & -5.566E-2 &  4.821E-4 &  2.874E-4 &  1.098E-4 
                      &  0        &  0        &  0        & $<$0.0001  \\
$^{3}P_{1}$           & -5.479E-2 & -5.916E-4 &  8.636E-4 & -2.065E-4 
	              & -2.721E-4 &  0        &  0        &  0.0002    \\
$^{3}P_{2}$           & -1.294E-2 &  3.075E-4 &  7.469E-4 & -4.289E-4 
                      & -2.179E-4 &  0        &  0        &  0.0002    \\
%$^{3}P_{2}-^{3}F_{2}$ & -3.953E-5 &  2.557E-5 & -1.371E-5 & -1.133E-5 
%                      &  3.740E-6 &  2.058E-6 & -7.076E-7 & $<$0.0001  \\
%$^{3}F_{2}$           & -1.610E-5 & -1.126E-6 &  7.666E-7 &  2.545E-7 
%                      & -1.356E-7 & -2.826E-8 & -2.888E-8             \\

\end{tabular}
\end{table*}

As mentioned earlier, the counter term coefficients are determined 
using standard fitting techniques.  We perform this fitting over all partial
wave channels, and find consistently very good agreement. In Fig. 1 
we compare some $^1S_0$ and $^3S_1$ matrix elements of 
$(PV_{bare}P+V_{CT})$ 
with those of $V_{low-k}$ for momenta below the cutoff $\Lambda$. 
A similar comparison for the $^3S_1-$$^3D_1$ and $^3D_1$ channels is 
displayed in Fig. 2.  We have also obtained very good agreement for the
P-waves, as displayed in Fig. 3. It may be mentioned that here
the momentum factor $qq'$ of Eq.(6) is essential 
in achieving the  good fit as shown.
As demonstrated by these figures, the two methods yield nearly identical 
effective interactions, lending strong support that 
$V_{low-k}$ can be very accurately represented by (P$V_{bare}$P+$V_{CT}$).
Furthermore $V_{CT}$
is a very short ranged effective interaction.

To further check the accuracy of our counter term approach,
we have calculated the phase shifts given by $V_{low-k}$
and compared them with those given by ($V_{bare}+V_{CT}$), as seen in Fig 4, 
where we have found that the phase shifts are almost exactly preserved by
the counter term approach. The phase shifts are
plotted up to a lab energy of 325 MeV, and only at this high momentum, do 
they begin to differ slightly from those obtained from $V_{low-k}$.
In passing, we mention that the deuteron binding energy is also accurately
reproduced by our counter term approach (e.g., for the CD Bonn
potential the deuteron binding energy given by the full potential versus
the counter term approach are respectively 2.225 and 2.224 MeV).
Thus, we can accurately reproduce $V_{low-k}$ and all its results by using 
($PV_{bare}P + V_{CT}$).

% We note that in the present work the counter term coefficients 
%are determined from the Hermitian $V_{low-k}$, which is obtained 
%by performing an Okubo transformation mentioned earlier.
%We have verified that the deuteron binding energy and low-energy
%phase shifts are also preserved by the Hermitian $V_{low-k}$.
%Since the non-Hermiticity of the ALS interaction is very slight,
%the counter terms obtained for these two interactions have been found to 
%be quite similar to each other.

\begin{table}
\caption{Comparison  of counter terms for $V_{low-k}$
obtained from the  CD-Bonn, Argonne V18, Nijmegen, and  Paris potentials using 
$\Lambda = \rm{2fm^{-1}}$. The units for the counter coefficients
are the same as in Table I. }
\vskip 0.5cm
\begin{tabular}{cccccc}
%\begin{tabular}
%{|cp{0.5in}|cp{0.5in}|cp{0.5in}|cp{1.0in}|cp{1.0in}|cp{1.0in}}
\hline
                & $C_0$  &  $C_2 $ & $C_4$   &    $C_4'$ &             \\
\hline
$^1S_0  $       & -0.158 & -0.0131 &  0.0004 & -0.0008   & CDB     \\ 
                & -0.570 &  0.0111 & -0.0005 &  0.0004   & V18  \\
                & -0.753 & -0.0099 &  0.0003 &  0.0002   & NIJ  \\ 
                & -1.162 & -0.0187 &  0.0004 & -0.0002   & PAR  \\ 

$ ^3S_1$        & -0.465 &  0.0588 & -0.0018 & -0.0116   &          \\
 	        & -1.081 &  0.0822 & -0.0002 & -0.0107   &          \\ 
                & -1.147 &  0.0682 &  0.0004 & -0.0100   &          \\ 
                & -2.228 &  0.0251 &  0.0013 & -0.0103   &          \\ 

$^3S_1-$$^3D_1$ & 0.0288 & -0.0036 &  0.0020 & -0.0037   &          \\
	        & 0.0209 & -0.0025 &  0.0018 & -0.0037   &          \\
                & 0.0242 & -0.0006 &  0.0040 & -0.0094   &          \\ 
                & 0.0184 & -0.0015 &  0.0019 & -0.0061   &          \\ 

%$ ^3D_1 $     &  0     & -0.0005 & -0.0005 &  & \\
% 	      &  0     & -0.0006 & -0.0006 &  & \\ 
%              &  0     & -0.0005 & -0.0005 &  & \\ 
%              &  0     & -0.0004 & -0.0004 &  & \\ 
\end{tabular}
\end{table}

Now let us examine the counter terms themselves.  In Table I, we list some 
of the counter term coefficients, using CD-Bonn as our bare potential. 
It is seen that the counter terms are significant only for the
S, P and D partial waves. 
We do not list results beyond the $^3P_2$ partial waves, as the coefficients
for them are all zero up to the level of our numerical accuracy (all entries 
in the table with magnitudes less than $10^{-4}$ have been set to zero).
It is of interest that except for 
the above partial waves, $V_{low-k}$ is essentially the same as
$PV_{bare}P$ alone. This behavior is clearly a reflection that
$V_{CT}$ is basically a very short ranged effective interaction.
From the table, $C_0$ is clearly the dominant term in the expansion.
Coefficients beyond $C_4$ are generally small and can be ignored.
 In the last
row of the table, we list the rms deviations between $V_{low-k}$ and 
$PV_{bare}P+V_{CT}$; the fit is indeed very good.
 
Comparing counter term coefficients for different $V_{bare}$ potentials 
can illustrate key differences between those potentials. 
Thus, in Table II, we 
compare the low order counter terms obtained for the CD-Bonn 
\cite{cdbonn}, 
Nijmegen \cite{nijmegen}, Argonne \cite{argonne} and Paris \cite{paris} NN 
potentials. The $C_0$ coefficients for these potentials are significantly 
different, indicating that one of the chief differences between these potentials
is the way in which they treat the short range repulsion. For instance, the 
Paris potential effectively has a very strong short-range repulsion and 
consequently its $C_0$ is of much larger magnitude than the others.

 Our counter term $V_{CT}$ has been determined by requiring a best fit
between ($V_{bare}$+$V_{CT}$) and $V_{low-k}$. We have explored other
schemes of fitting: We have tried to determine $V_{CT}$ by requiring
a best fit  between ($V_{one-pion}$+$V_{CT}$) and $V_{low-k}$.
But the results are far from satisfactory, the resulting rms deviation
at best fit being too large. It may be of interest to determine 
$V_{CT}$ by requiring
a best fit between $V_{low-k}$ and $V_{CT}$ alone. We have also tried this,
with similar unsatisfactory results (large resulting rms deviation).
A possible explanation may be the following. A main portion of
$V_{low-k}$ come from $V_{two-pion}$ and other higher order
processes; it appears that such contributions can not be compensated
by the $V_{CT}$ of the simple low-order form of Eq.(6).

\section{Scaling of counter terms}

 In Fig. 5 we display the scaling behavior of counter terms $C_0$ and $C_2$
with respect to the decimation momentum $\Lambda$. We note that for the 
$^1S_0$ channel, $C_0$ and $C_2$ display a weak $\Lambda$ dependence. This is
a welcome result. We should note that in MEEFT, we are obliged to choose 
$\Lambda=\rm2.1fm^{-1}$, because phase shifts from experiments have been 
carried out up through this momentum and the {\it more effective} in MEEFT
means that we fit all available experimental data by $V_{low-k}$.
However, the $C_0$ for the $^3S_1$ channel varies
significantly with $\Lambda$. In this channel, there is tensor force
which is a mid-range interaction coming from $\pi$ and $\rho$ mesons;
it has large  momentum components
in the intermediate momentum region of several $\rm fm^{-1}$.
As we lower $\Lambda$ through this region, we are actually integrating out 
a predominant portion of the tensor force, in addition to
the short range repulsion of $V_{bare}$. As a result, we would expect 
the $C_0$ to change to compensate for this loss of the tensor force. 
In contrast, the $C_0$ for $^1S_0$ comes mainly
from integrating out only the short range repulsion, and little variation
is seen. Conversely, as $\Lambda$ 
increases, the tensor force is largely retained within the cutoff, and the 
two $C_0$'s come close to each other as seen in the figure. To 
justify this line of thinking, we removed the tensor force to see
how it would effect the $^3S_1$ $C_0$.  We found that the scaling for this term
was indeed quite different, displaying a flat behavior over the entire $\Lambda$
range; $C_0$ only changed from 0.05 at $\Lambda=1.2$fm to 0.07 at 
$\Lambda=3.0$fm.

\begin{figure}
\rotatebox{270}{\scalebox{0.4}{\includegraphics{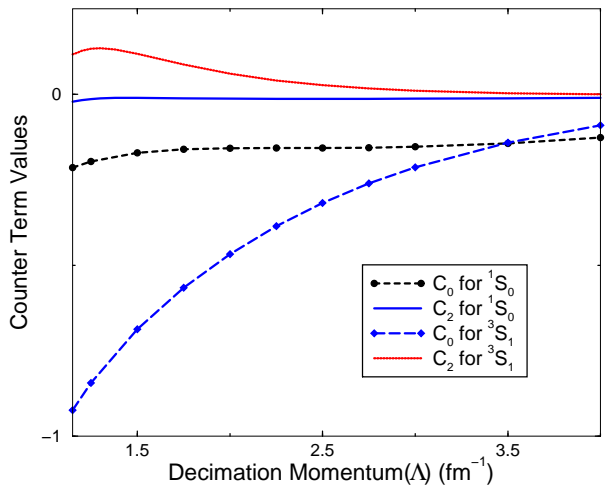}}}
%{\epsfig{file=ctholtfig2.eps,width=14cm,angle=270}}
\caption{Scaling behavior of S-wave leading order counter terms.}
\label{fig.5}
\end{figure}

It should be observed that our Fig. 5 does not cover the small $\Lambda$ region.
The reason for this is that we have found that our counter term approach is not 
applicable to
cases with $\Lambda \sim 1 fm^{-1}$ or smaller. In this region, we have not been 
able to achieve a satisfactory fit between $V_{low-k}$ and ($V_{bare}+
V_{CT}$). A likely reason for this is that for small $\Lambda$'s,
we are integrating out a major part of the tensor force, which can not
be compensated by the simple low order counter terms as indicated by
Eq.(6).

As we mentioned earlier, the counter terms are all rather small except
for the S waves. This is consistent with the RG-EFT approach where the 
counter term potential is mainly  a delta function force \cite{lepage,epel99}. 
When using $\Lambda \sim 2 fm^{-1}$, we see that the $C_0$ coefficients for the 
$^1S_0$ and $^3S_1$ 
channels are different. This suggests that the counter term 
potential is spin dependent and may be written as
\begin{equation}
 V_{CT}\approx (f_0 +f'_0\sigma \cdot \sigma)
\delta (\vec r),
\end{equation}
 with $f_0=(3C_0(^3S_1)+C_0(^1S_0))/4$ and
$f'_0=(C_0(^3S_1)-C_0(^1S_0))/4$. For the coefficients of Table I, we have
$f_0=-0.382$ fm and $f'_0=-0.077$ fm. As indicated in Fig. 5, this 
spin-dependent factor $f'_0$ will diminish as $\Lambda$ increases.

\section{Comparison of shell model matrix elements}

In shell model calculations, a basic input is the relative matrix elements 
$\langle nlSJT \mid V_{eff} \mid n'l'SJT \rangle$, where $nl$ and $n'l'$ 
denote the oscillator wave functions for the relative motion of the two 
nucleons and 
$S$ and $T$ are the two-nucleon spin and isospin. The relative momenta 
$l(l')$ and $S$ couple to total angular momentum $J$. From these matrix 
elements, the two-particle shell model matrix elements in the laboratory 
frame
can be calculated \cite{kuobrown}. Thus if $V_{low-k}$ and 
$PV_{bare}P+V_{CT}$
are to prove useful in a broad sense, they must preserve these matrix 
elements.

\begin{table}
\caption{Shell model relative matrix elements of $V_{low-k}$
 and ($PV_{NN}P+V_{CT}$) (column CT), with $\Lambda$
= 2.0 $\rm{fm^{-1}}$ and $\hbar \omega  =$19 MeV. Matrix elements are in units
of MeV. }
\vskip 0.5cm
\begin{tabular}
{cp{0.5in}cp{0.5in}cp{0.5in}cp{1.0in}cp{1.0in}cp{1.0in}cp{1.0in}}
\hline
$TSj$& $nl$& $n'l'$&   CDB  & CDB-CT &   Nij   &  V18   \\
\hline
100  & 00  & 00  & -9.053 & -9.053 & -9.066  & -9.062 \\ 
     & 10  & 00  & -6.744 & -6.742 & -6.698  & -6.609 \\
     & 10  & 20  & -3.527 & -3.529 & -3.363  & -3.375 \\
011  & 00  & 00  & -11.00 & -11.00 & -10.72  & -10.80 \\
     & 02  & 00  & -8.440 & -8.405 & -8.636  & -8.627 \\
     & 02  & 02  &  2.450 &  2.453 &  2.253  &  2.263 \\
     & 10  & 20  & -6.873 & -6.850 & -7.033  & -6.738 \\
     & 10  & 22  & -6.969 & -6.940 & -7.562  & -7.637 \\
     & 12  & 22  &  2.281 &  2.278 &  2.070  &  2.079 \\
001  & 01  & 01  &  3.835 &  3.858 &  3.587  &  3.670 \\
     & 11  & 01  &  3.757 &  3.781 &  3.446  &  3.440 \\
110  & 01  & 01  & -2.482 & -2.452 & -2.407  & -2.557 \\
     & 11  & 01  & -1.124 & -1.099 & -0.900  & -1.162 \\
\end{tabular}
\end{table}

In Table III we provide a comparison between these matrix elements as 
obtained 
from several input potentials.  To begin, we note that the harmonic 
oscillator matrix elements for $V_{low-k}$ are approximately the same 
regardless of which of the four bare potentials we use.  So, as far as 
providing input for shell model calculations, this is further evidence 
that 
the $V_{low-k}$ is approximately unique.  Next, we come to the main result 
of 
the table, which is that the matrix elements for $V_{low-k}$ and 
$V_{bare}+V_{CT}$ are in very good agreement.  We have used the CD Bonn 
for 
the bare potential, but this holds for the other bare potentials as well.  
In 
fact for the $^{1}S_{0}$, $^{3}S_{1}$, and $^{3}D_{1}$ channels, the 
elements 
are virtually exact out to four significant figures, and there is only 
minimal disagreement in the coupled channel.

\section{Conclusion}

While there are several models for the nucleon-nucleon potential in use today,
they suffer from uncertainty and model dependence. Motivated by RG-EFT ideas,
a low momentum NN interaction, $V_{low-k}$, has been constructed via 
integrating out the high momentum, model dependent components of these 
different potentials. The result appears to give an approximately unique 
representation of the low momentum NN potential, and the main issue addressed 
in this paper was whether $V_{low-k}$ could accurately be cast in a form
$V_{bare}+V_{CT}$, where $V_{CT}$ is a low order counter term series, and what
the physical significance of this counter term series was.  We have 
shown that this was indeed the case as $V_{low-k}$ is nearly identical to 
$V_{NN}+V_{CT}$ over all partial waves, and that $V_{NN}+V_{CT}$ 
reproduces both the deuteron binding energy and the NN
phase shifts in the low momentum region. We have found that only the leading 
terms in the counter term series have much significance, indicating that the
counter term potential is mainly a short range effective interaction
which can be accurately represented by a simple low-order momentum expansion.
Furthermore, we examined the
scaling properties of the S-wave counter terms with respect to $\Lambda$
and found that the tensor force is responsible for the behavioral differences
exhibited between the $^3S_1$ channel and the $^1S_0$ channel. 
Finally, we  examined the potential for using $V_{low-k}$ and 
$V_{NN}+V_{CT}$ in shell model calculations.  Again, it was seen that not 
only
does $V_{low-k}$ give approximately identical input matrix elements 
irrespective of which bare potential was used, but that the counter term 
approach provided nearly the same results.  Thus we have shown that the 
high
momentum information integrated out of each bare NN potential can be 
accurately
replaced by the counter terms, making the use of $V_{low-k}$ in a broad 
context much more tractable.

\begin{acknowledgments} 
Many helpful discussions with Rupreht Machleidt and Achim Schwenk are 
gratefully acknowledged.
This work was supported in part by the U.S. DOE Grant No.  DE-FG02-88ER40388. 
\end{acknowledgments}

%\newpage

%\centerline{\bf Appendix}
%\vskip 0.5cm

%\newpage

\end{document}